# On-Sky Tests of an A/R Coated Silicon Grism on board NICS@TNG

F. Vitali*[a], V. Foglietti[b], D. Lorenzetti[a], E. Cianci[c], F. Ghinassi[d], A. Harutyunyan[d], S. Antoniucci[a], C. Riverol[d], L. Riverol[d]

[a]INAF - Osservatorio Astronomico di Roma, Via Frascati 33, 00040-I Monte Porzio Catone - Italy
[b]CNR - Istituto di Struttura della Materia, Area della Ricerca Roma 1, Montelibretti, Via Salaria, Km. 29,300, 00016-I Monterotondo - Italy
[c]CNR- Istituto per la Microelettronica e Microsistemi, Laboratorio MDM, Via C. Olivetti 2, 20864-I Agrate Brianza (MB) - Italy
[d]Fundación Galileo Galilei - INAF, Rambla José Ana Fernández Pérez, 7, 38712 Breña Baja, TF - Spain

## ABSTRACT

We present the results of our project for the design and construction and on-sky test of silicon grisms. The fabrication of such devices is a complex and critical process involving litho-masking, anisotropic etching and direct bonding techniques. After the successful fabrication of the silicon grating, we have optimized the bonding of the grating onto the hypotenuse of a silicon prism to get the final prototype. After some critical phases during the experimentation a silicon grism with 363 grooves/mm and a blaze angle of 14 degrees has been eventually fabricated. The application of an A/R coating on both the surfaces has been the last step: this procedure is critical because of the groove geometry of the diffraction grating, whose performace might be compromised by the coating. Then, the grism was inserted in the filter wheel of the Near Infrared camera NICS, at the focal plane of the National Galileo Telescope (TNG), the 3.5 m Italian facility in the Canary Islands (E). The result of the on-sky tests are given in detail.

## 1. INTRODUCTION

We have ultimated a long term project on the construction of a silicon grism and performed on-sky tests to derive the main optical response of the device.

The production of a silicon grism is a complex procedure, involving nanotechnologies processes, due to the cristalline nature of the silicon. In few words, the process consists in e-beam lithographic mask replication, anisotropic etching and the grating-prism bonding. Our entire construction process has been detailed reported in Vitali *et al.,* 2000[1], Cianci *et al.,* 2001[2], Vitali *et al.,* 2003[3] and Vitali *et al.,* 2008[4].

The produced prototype is currently mounted in the near infrared camera NICS[5], on board at the TNG telescope (Canary Islands, E). The main characteristics of the grism are reported in Table 1.

*fabrizio.vitali@oa-roma.inaf.it; phone +39 06 94286462; fax +39 06 9447243

After the completion of the silicon grism, we performed the very last step, i.e. the application of an A/R coating on both surfaces of the grism, a critical process, in which the grooved surface of the grating could be altered by the coating process. We finally take advantage of the succesful experience done by Gully-Santiago and collaborators (Gully-Santiago *et al.,* 2010[6]) with the II-VI optical firm.

Table 1. The main characteristics of the produced silicon grism.

| Grating Size | 32 x 32 mm |
|---|---|
| Groove density | 363.6 gr/mm |
| Grating Pitch | 2.75 µm |
| Blaze Angle | 14 deg |
| Flat Top Size | < 0.2 µm |
| RMS roughness | < 1.0 nm |
| Prism Angle | 14 deg |
| Undeviated Wavelength | 1.64 µm |

After the A/R coating has been applied, we had the complete prototype and we eventually put it in the grism wheel of the NICS imager-spectrometer at the Italian National Telescope Galileo (TNG), in the Canary Islands, during April 2013. A Science verification phase has been performed during three nights on May, 2013: the technical and scientific resulta are reported herein detail.

## 2. THE LAST STEP: THE A/R COATING

The application of an A/R coating on the grating surface is a critical process because it could modify or damage the grooves geoometry. Moreover, we were worried about the prism-grating interface, that during the vacuum and high temperature coating process could break and destroy the grism.

The recent experience of Gully-Santiago and collaborators[6] , together with II-VI optical firm, has shown that a proper coating can be applied also onto the grating surface, without modifying the optical performances of a monolithic grism. This makes possible to reduce the Fresnel losses, due to the high index of refraction of the silicon, that can result in about 30% per surface. The II-VI optical firm offered an A/R coating with reflection and transmission sample curves shown in Figure 1 and we decided to apply it to our grism. The coating process returned us a safe grism (Figure 2), with very good performances in terms of surfaces reflection and transmission (Figure 3), with a peak of efficiency of about 72% at $\lambda$=1.73 µm.

### 2.1 The weak side of our silicon grism

We prepared two grisms to send to II-VI for the A/R coating application. During the interaction with the firm, they asked to remove a layer of red silicon between the prism and the grating, that we had put to avoid dust or other pollutions between the two parts f the grism. This silicon could possibly pollute the vacuum coating chamber, then we tried to remove it. Unfortunately, during this operation, one of the two gratings detached from the prism and we were not able to bond it again.

This means that the bonding operation is really critical and can not be controlled perfectly. This led us to focus our future efforts in producing monolithic grisms, avoiding the e-beam lithography and writing the grating directly on the prism face. To this aim, the current optical projection lithography systems are appropriate and with an i-line ( 365 nm) stepper the required resoluton can be easily obtained as well.

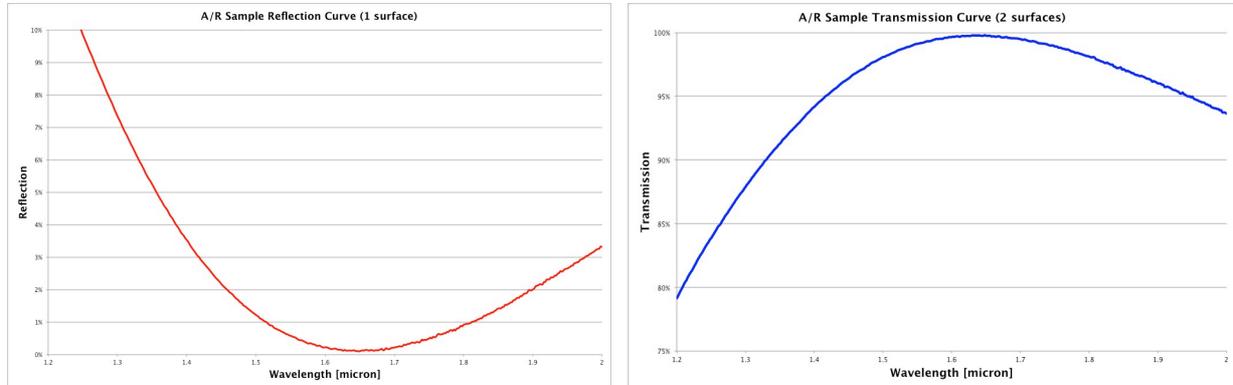

Figure 1. The reflection (left) and trasmission (right) A/R sample curves from II-VI.

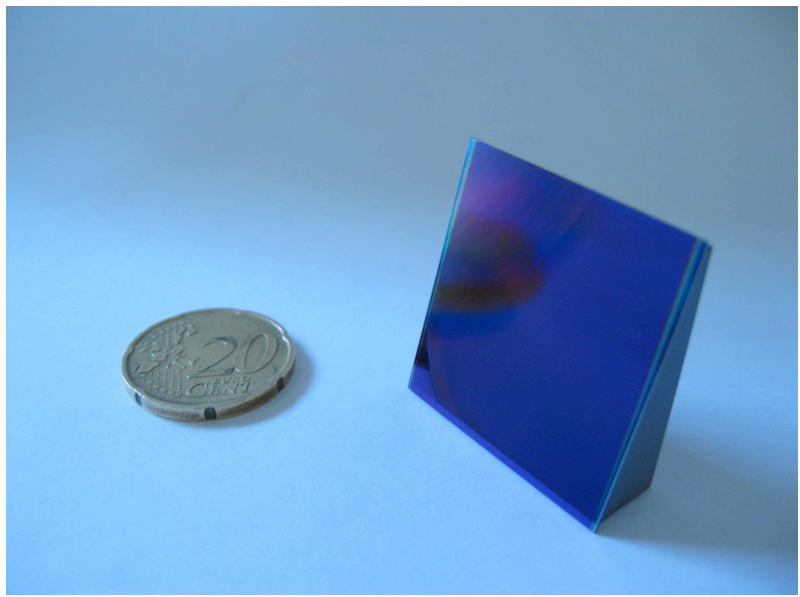

Figure 2. The silicon grism, as delivered by II-VI.

## 3. THE SILICON GRISM ONBOARD NICS@TNG

The silicon grism has been finally inserted in the grism wheel (Figure 4, left) of the near infrared spectro-imager camera NICS @ TNG, and dubbed SiGH (SIlicon Grism in the H band), during a campaign in April, 2013. The grism has been mounted with a simple aluminum holder and tightened with four metal clamps at each angles of the grism (Figure 4, right). To avoid any damage to the grating surface, the clamps have been covered with shrink tubes (Thermofit).

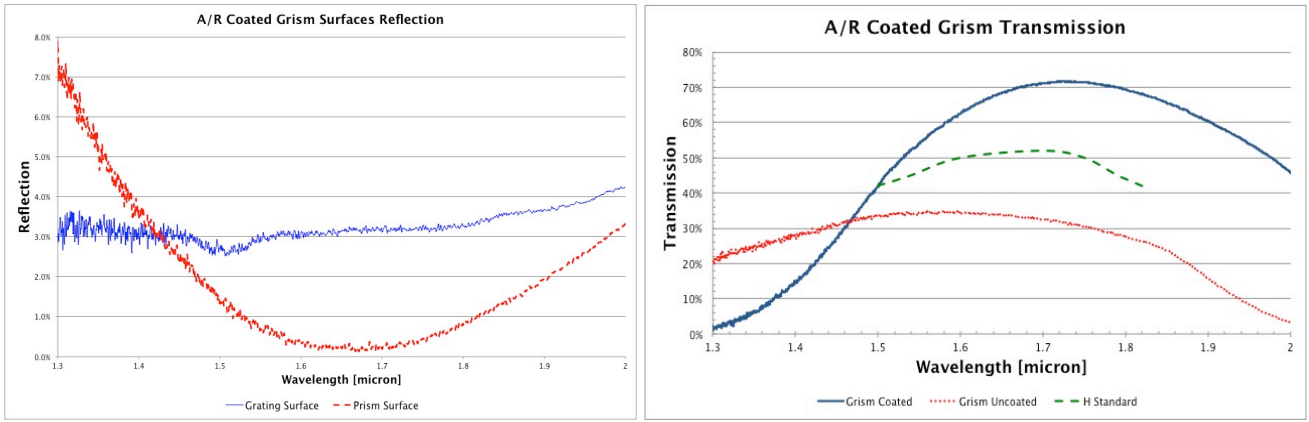

Figure 3. Left: the grism surfaces reflection as measured by II-VI. The grating surface (blue) has an average reflection of about 3% in the H band. Right: the transmission curve before (dotted red) and after (solid blue) the A/R coating application. The efficiency of the H standard resina replica grism (dashed green) already hosted in the camera-spectrometer NICS is shown as well for comparison.

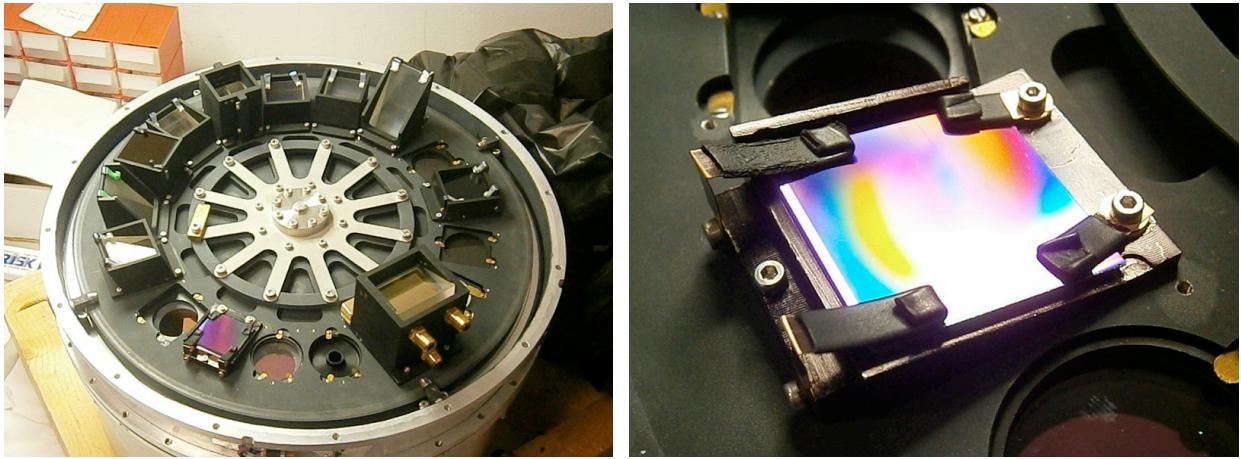

Figure 4. SiGH mounted in the grism wheel of NICS (left) and a detail of the holding clamps (right)

## 4. THE ON-SKY TESTS AT TNG

The performances of the silicon grism have been checked at the telescope during a Science Verification Phase during May, 2013. We firstly checked the technical response of the grism, then some science targets were observed to compare the spectra with some already obtained with other telescope and spectrometers.

## 4.1 The performances of SiGH

The first light of SiGH has been made on a calibration Argon lamp, that produced the spectrum image shown in Figure 5. The main lines are well defined and visible in the range 1.5 - 1.75 µm (the 50% cut of the H filter is at 1.48 and 1.78 µm).

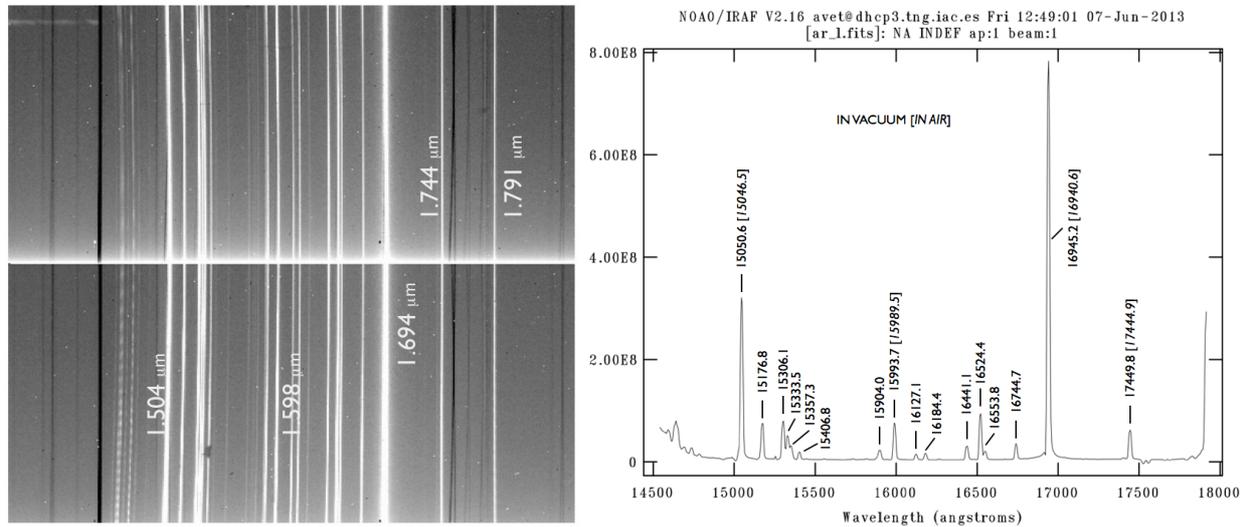

Figure 5. The spectral image of the Argon calibration lamp (left) and the extracted spectrum with the main lines wavelength labeled, as measured in air and in vacuum.

In Figure 6, we have compared the spectrum of the standard star AS34 taken with SiGH and with the H standard grism, already mounted in NICS, with about the same resolution. The figure shows clearly that SiGH is more efficient, as expected, over the whole range of the H band, with an average gain in efficiency of 1.25.

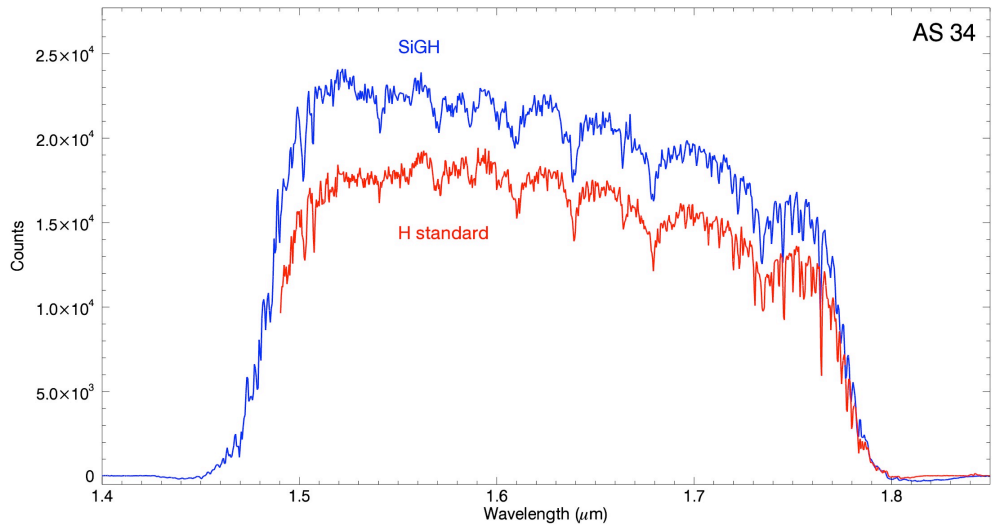

Figure 6. The standard star AS34 observed with SiGH (red, above) and the H standard grism (blue, below), already mounted in NICS.

In Figure 7 we show the spectrum of the source HBC722 taken with different slits, 1" and 0.5". With the former, SiGH provides a resolution of about 1.000, as expected.

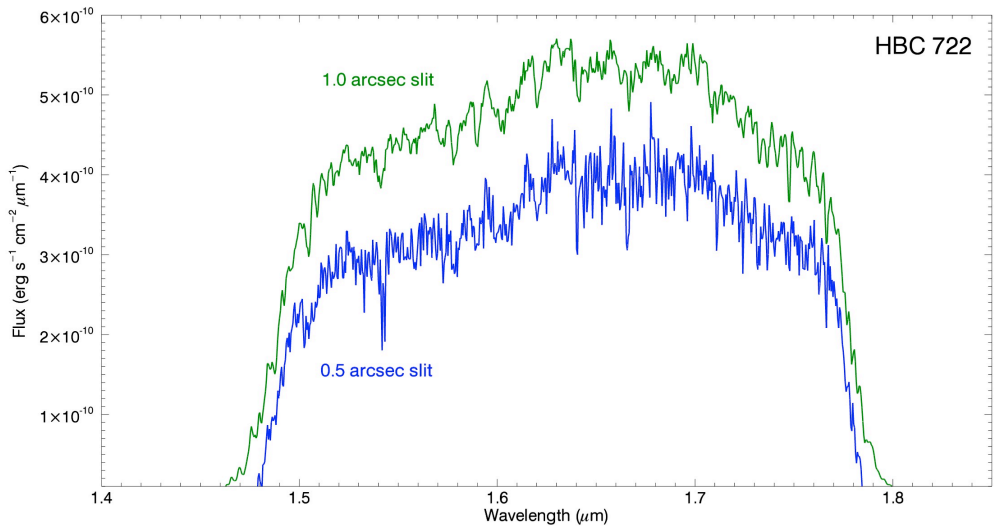

Figure 7. The source HBC722 observed with two different slits, 1" (green, above) and 0.5" (blue, below), to compare the relative resolution. The 0.5" spectrum has been decreased of $1\times10^{-10}$ in flux to be easily compared.

### 4.2 The science targets

The observations in one single band (H in this case) can not provide relevant scientific informations. In any case, we would like to test the response of our grism, compared with other observations we have collected so far with other instruments and telescope. This can give us precious information about the reliability of our grism.

We selected some Young Stellar Objects (YSOs) from our EXORCISM[7] catalogue, ideally suited to test the performances of our SiGH, since in their near-IR spectra are present many bright atomic emission lines, originating in the active regions of their circumstellar environment (i.e. the accretion disk and the associated winds/jets). EXORCISM project aims to observe EXOr stars in differrent phases of their activity, both in bursts and quiescience: EXors are pre-main sequence eruptive stars showing intermittent outbursts (Δmag about 4- 5) of relative short duration (months), superposed to longer (years) quiescence periods. These bursts, usually detected in the optical and near-IR bands, are related to disk accretion events in which there is a sudden increase of the mass accretion rate by orders of magnitude (e.g. Hartmann & Kenyon 1996[8]).

In Figure 8 we show the spectrum of IRS54 taken with SiGH at NICS-TNG with the ones taken in 2005 with ISAAC-VLT, with a much higher resolution, of about 10.000. The source is showing a wide variety of emission features dominated by HI recombination lines and currently is undergoing an outburst phase, as can be easily seen from the flux level of our spectrum. Moreover, as already observed with previous photometric observation, also the slope of the spectrum is changed, possibly correlated with some mechanism strongly related to the accretion onto the central source.

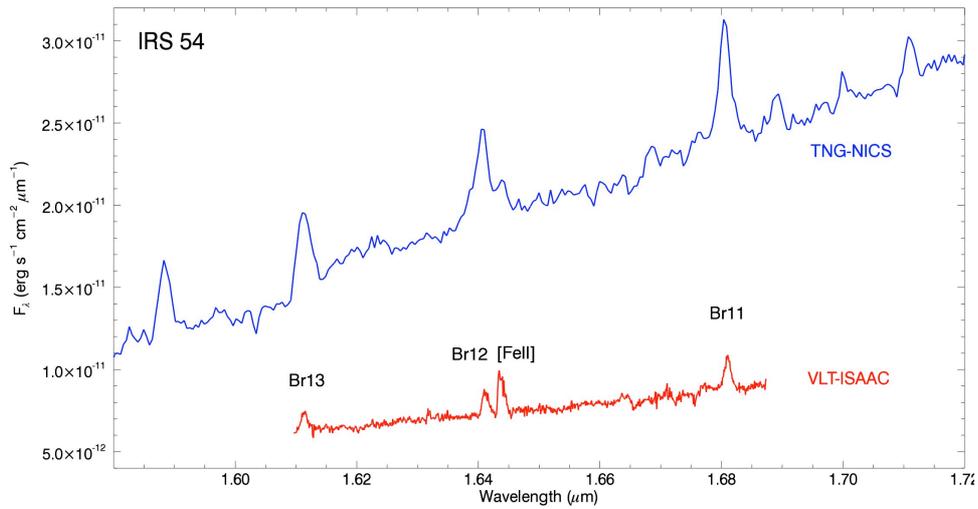

Figure 8. The spectrum of IRS54, taken with SiGH at NICS-TNG (R~1.000), compared with that taken at ISAAC-VLT (R~10.000) in 2005. The lines of the Brackett series are clearly visible and currently the source clearly exhibit an eruptive phase. The spectrum is higher in flux and changed the slope, as expected.

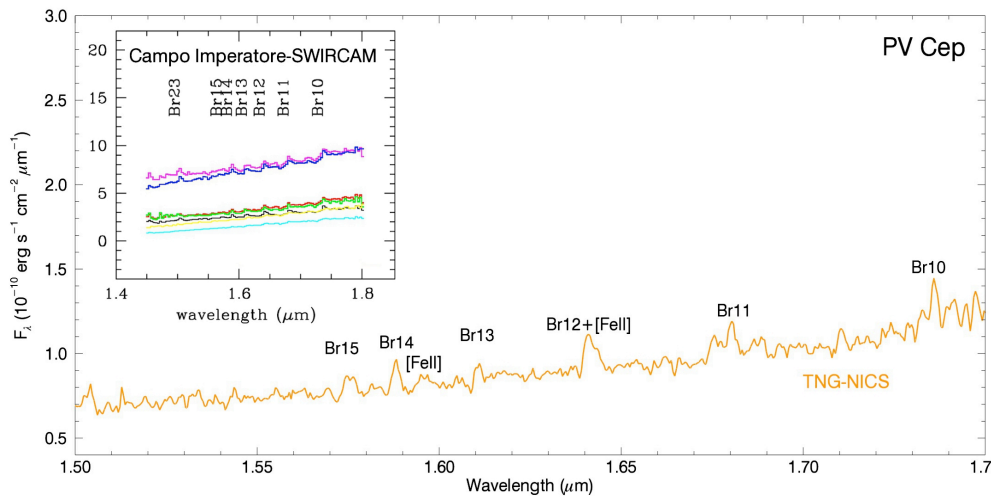

Figure 9. The spectrum of PV Cep taken with SiGH, compared with the same taken in 2007 with SWIRCAM-Campo Imperatore (AQ, Italy), with R~100. The flux scale are homogeneous, then the source is now in a quiescent phase.

## 5. CONCLUSIONS

We have presented the final results of our experimentation about the construction of a silicon grism. The final prototype, complete of an A/R coating on both surfaces, has been produced and mounted inside the imager-spectrometer NICS at the TNG telescope. During the last phase, the A/R coating application, we found that the bonding of the grating onto the prism surface is a very critical step, that led us to focus our next effort on the construction of a monolithic grism

The efficiency of the grism is about 72% at 1.7 μm, comparable with other similar devices, available in other instruments. The working wavelength range and resolution are fully in accordance with the expectations. The on-sky test at the TNG telescope provided valid scientific frames, confirming the reliability of the device.

## ACKNOWLEDGEMENTS

The author would like to thanks Dr. Gully-santiago for his kind availability. Moreover, the author wish to thank the Director of TNG, Dr. Emilio Molinari and all the staff at TNG for their kind hospitality.

The author would like also thank Dr. J. Bacon and Dr. Brando of II-VI fror their valuable collaboration during the A/R coating phase. This last part of the project on the silicon grism was supported by INAF with a National grant TECNO-PRIN, Prot. 4387/2010.